\input{aipcheck}

\documentclass[
    ,final            % use final for the camera ready runs
  ]
  {aipproc}

\layoutstyle{6x9}

\def\ltsima{$\; \buildrel < \over \sim \;$}
\def\lsim{\lower.5ex\hbox{\ltsima}}
\def\gtsima{$\; \buildrel > \over \sim \;$}
\def\gsim{\lower.5ex\hbox{\gtsima}}

\newcommand{\be}{\begin{equation}}
\newcommand{\en}{\end{equation}}
\newcommand{\ergs}{\rm \ erg \; s^{-1}}

\begin{document}

\title{Emission processes in quiescent neutron star transients}

\author{Sergio Campana}{
  address={INAF -- Osservatorio astronomico di Brera, Via Bianchi 46, 23807
Merate (LC), Italy}
}

\begin{abstract}
We review the observational properties of transient systems made by a neutron star
primary and a late dwarf companion (known also as Soft X--ray Transients) during their 
quiescent state. We focus on the several emission mechanisms proposed and 
try to compare them with observations. Finally, we review new tools to improve
our comprehension of the physics of the emission processes. 
\end{abstract}

\maketitle

%%%%%%%%%%%%%%%%%%%%%%%%%%%%%%%%%%%%%%%%%%%%
%% MAINMATTER
%%%%%%%%%%%%%%%%%%%%%%%%%%%%%%%%%%%%%%%%%%%%

\section{What are neutron star transients?}

Neutron star transients (also known as Soft X--ray transients, SXRTs) are
binary systems with a late star companion and a neutron star primary. Orbital
periods are short (less than a day) and the companion fills its Roche lobe
transferring matter to the primary through the first Lagrangian point.
Outflowing matter has a large angular momentum and falling onto the primary
forms an accretion disk, which mediates the mass transfer.
Despite persistent sources, known as Low Mass X--ray Binaries (LMXRBs), SXRTs
alternate periods of quiescence, during which they attain an X--ray luminosity of
$L_X\sim 10^{32}-10^{33}\ergs$ to periods (lasting weeks to months) during which they
are as bright as their relatives (i.e. LMXRBs, $L_X\sim
10^{36}-10^{38}\ergs$). Actually, during these bright periods (called
outbursts) SXRTs share all the same characteristics of LMXRBs.  Recurrence
times vary from $\sim 1$ to $>30$ years. For a review see Campana et al. (1998a).

\section{Why do we study transients?}

Transient sources vary their luminosity over several orders of
magnitude. These variations reflect, at least partially, in variations of the
mass inflow rate toward the compact object and allow to sample a variety of
physical conditions and regimes that are inaccessible to persistent
(bright) sources. Especially interesting are SXRTs in quiescence when X--ray
emission can in principle be powered by mechanisms not involving the inflow of
matter onto the neutron star surface.

\section{Mean quiescent X--ray luminosity and spectra}

The study of SXRTs in quiescence is hampered by their low luminosity. A
handful of sources were known before Chandra and XMM-Newton.
In particular, Chandra is discovering a large number of SXRTs in globular
clusters\footnote{Actually there is a bias in these discoveries since sources
are preferentially selected if they are `bright' ($L_X\sim 10^{33}\ergs$) and
with a soft spectrum. Recent discoveries have shown that SXRTs may also show 
only a hard power law tail (SAX J1808.4--3658, Campana et al. 2002; EXO
1745-248 Wijnands et al. 2003). These sources are more difficult to pinpoint due to
their similarities with cataclysmic variables.} ($\sim 15$,
e.g. Pooley et al. 2003), whereas XMM-Newton is providing good spectral
information for a sizable number of sources. For all the sources discovered so
far the quiescent 0.5--10 keV luminosity is in the range $10^{32} - {\rm few}
10^{33}\ergs$. Only one source is outside this range and it is the first
discovered millisecond transient X--ray pulsar SAX J1808.4--3658
(Wijnands \& van der Klis 1998). This source has been studied through XMM-Newton
observations indicating a quiescent luminosity of $5\times10^{31}\ergs$
(Campana et al. 2002).  

On the spectral side, common behaviors can also been found. Quiescent X--ray
spectra of SXRTs are usually characterized by two spectral components: 1) a soft
component modeled as a black body or, more physically, by cooling emission
from the entire neutron star surface; 2) a hard power law energy tail.
The first component comprises the majority of the flux ($50-100\%$). The power
law tail is present only in a fraction of sources and contributes up to $50\%$
in the 0.5--10 keV energy band (e.g. Campana 2001). 

%\begin{figure}
%  \includegraphics[height=.3\textheight]{golfer}
%  \caption{Picture to fixed height}
%\end{figure}

\subsection{Similarities and differences with black hole transients}

Luminosities and spectra of neutron star transients are markedly different
from the ones observed in quiescent transients containing a black hole (TBH)
as primary. In the case of TBH the 0.5--10 keV luminosities range between
$10^{30}-10^{31}\ergs$ (i.e. a factor of $\sim 100$ below SXRTs) and their faint
quiescent spectra show indication for the presence of only a power law tail
(Garcia et al. 2001, Kong et al. 2002).

\section{Models for the quiescent emission}

Several models for the quiescent emission of SXRTs have been put
forward. These range from accretion of matter onto the neutron star, in
several flavors (advection-dominated, convection-dominated, with outflows,
etc., Menou et al. 1999, Menou et al. 2001) to jets (Fender et al. 2003), 
to neutron star cooling after long-term ($10^4$ yr) heating during outbursts
(Brown et al. 1998), to emission regimes connected to the presence of a
magnetic field (propeller and radio pulsar turn on,
Campana et al. 1998b, Campana \& Stella 2000). 

\begin{itemize}
\item Advection-dominated accretion flow (ADAF) models should naturally predict the
lower X--ray luminosity observed in TBHs with respect to SXRTs (since the
innermost accretion disk regions are advected inward into the black hole,
whereas for neutron stars the hard surface will release all the available
power). This is true only in principle and details fail to be accounted for
(in particular the luminosity ratio between neutron stars and black holes
quiescent luminosities of $\sim 100$, Menou et al. 1999).
Moreover, the ADAF expected spectrum for SXRTs can explain only the hard part
of the spectrum but not the soft component (Yu et al. 1996).
\item Jet emission has been recently proposed to account for the quiescent X--ray
emission of TBHs (Fender et al. 2003). This component however should be minor in
the case of neutron stars.
\item Cooling of the neutron star after deep crustal heating is the most popular
models to explain the soft component of quiescent SXRTs. Basically, the
neutron star emits black body-like radiation due to the heating of its interior
occurred during the repeated outbursts (Brown et al. 1998, Colpi et al. 2001). Fitting
the soft component with neutron star cooling models, several authors derived radii in
agreement with the expectations (i.e. $\sim 10-15$ km) and opening the
possibility of directly measuring the neutron star radii in sources of well
known distances (e.g. globular cluster sources). 
\item Regimes related to the presence of a neutron star magnetic field involve
the control of the neutron star magnetosphere of the motion of the incoming
matter. When the mass inflow is large the magnetosphere is compressed and
matter can reach the neutron star surface following the magnetic field lines. 
At lower mass inflow rates, the magnetosphere expands. For low enough rates,
the magnetosphere rotates (being anchored and corotating with the neutron
star) faster than the matter orbiting at Keplerian frequency around it. 
When matter tries to get attached to the field lines it experiments a centrifugal 
force larger than gravity and gets expelled. This is the propeller regime 
(Illarionov \& Sunyaev 1975). The accretion efficiency is reduced due to the fact 
that matter is stopped at the magnetosphere and does not reach the surface 
(thus releasing less potential energy). 

For even lower mass inflow rates, the magnetosphere starts rotating
at the speed of light. At this point the magnetic field lines open up and the
field changes from dipolar to radiative. In this situation we have a loss of
energy from the rotating neutron star according to the usual Hertz
formula (e.g. Campana et al. 1998a). This energy release induces a pressure on
the infalling matter sweeping it away. This should be the case for quiescent
SXRTs. The neutron star/radio pulsar relativistic wind interacts with matter
from the companion, generating a shock front in which a fraction ($\eta\sim
0.01-0.1$, depending mainly on geometry and weakly on mass inflow rate) 
of the total spin-down losses. The expected spectrum is a synchrotron one 
with photon index $\Gamma\sim 1.5-2$ (Campana et al. 1998a, Tavani \& Arons 1997). 
\end{itemize}

{\bf First summary.} Which emission models agree with the observational data of
SXRTs in quiescence? Actually there are two main models, even if there is no
consensus on them and are not particularly well defined. The first
consideration is that the soft and the hard components seem not to come from
the same emission mechanism. One class of models involve accretion onto the
neutron star surface as the main ingredient. The model presented in
Menou et al. (1999) involves an ADAF (responsible for the hard component) and a
propeller (for the soft component). Other possibilities, even if never
investigated in details, involve direct accretion onto the neutron star
surface (resulting in a soft spectrum, e.g. Zampieri et al. 1995) and a corona
for the power law.
The other family involves the cooling of the neutron star as responsible for
the soft component. The hard component is explained 
as the interaction of the pulsar relativistic wind with matter outflowing from
the companion (Campana et al. 1998a, Campana \& Stella2000).

\section{How can we improve the situation?}

\ \ {\bf Better X--ray spectra.} The first and easiest way is to obtain further
X--ray observations. But this will not directly produce breakthroughs unless
very peculiar sources, otherwise we will gain knowledge in the statistical
properties of the sources. The main problem is that the ADAF models do not
produce firm spectral predictions\footnote{This is almost true for the shock
emission scenario in which a synchrotron spectrum is expected but Compton
losses may steepen the spectrum.}.
 
{\bf Radio pulsar search.} In the shock emission scenario an active radio
pulsar is predicted. One can try to search for pulsed radio signals. These
have been searched for in a sample of 5 SXRTs in quiescence with negative
results (Burgay et al. 2003). This non-detection is significant since the
probability of having missed all the observed sources because of their
weakness or beaming is only about $25\%$.  However, as already noted
(Campana et al. 1998a, Burderi et al. 2003), radio emission can be severely hampered by
free-free absorption of matter around the system. High frequency searched
would be very valuable. 

{\bf X--ray variability.} A different path consists in studying well known
sources in much higher details than have been done since now. This is possible
thanks to the new X--ray facilities like Chandra and XMM-Newton and this has been
done on the two best studied SXRTs: Aql X-1 and Cen X-4.

$\bullet$ The first source has been monitored during quiescence with Chandra for 4 times
over 5 mouths with particular care to the systematic effects: Aql X-1 has
always been observed on the same position of the Chandra CCDs and with a
sub-imaging in order to reduce any pile-up effect (Rutledge et al. 2002). This
analysis showed that the temperature of the neutron star atmosphere varied as
$k\,T=130^{+3}_{-5}$ eV, down to $113^{+3}_{-4}$ eV, and finally increasing to
$118^{+9}_{-4}$ eV for the final two observations. A power law tail was
detected only in the last two observations. Short term variablity
($32^{+8}_{-6}\%$ rms) was also observed in the last observation (when the
power law tail contributes more). Given this variability in temperature the
cooling neutron star model is not able to explain the data and (Rutledge et
al. 2002) concluded that accretion onto the neutron star surface was more likely. 
Campana \& Stella (2003) approached the same data plus an unpublished BeppoSAX 
observation of Aql X-1 on the same epoch.
They obtained a good fit with a variable temperature model but a similarly
good fit for a varying power law plus column density model. These correlated
changes are expected based on the shock emission scenario in light of the
recent observations of the ms pulsar PSR 1740--5340 showing variable emission
along the orbit and with difference from orbit to orbit (D'Amico et
al. 2001). This testifies for a vary variable ambient around all these systems
which can result in correlated changes between the matter along the line of
sight and the energy emitted. 

$\bullet$ Cen X-4 was observed by XMM-Newton during quiescence. Given the closeness  
of Cen X-4 (1.2 kpc) this provides the observation of a SXRT in quiescence with the 
highest signal to noise ratio. The quiescent state of Cen X-4 has been recognised to
be variable both on long times scales ($\sim 40\%$ in 5 yr, Rutledge et al. 2001) 
and on shorter timescales (factor of $\sim 3$ in a few days, Campana et al. 1997). 
During this XMM-Newton observation X--ray variability has been observed
(at a level of $45\pm7\%$ rms in the $10^{-4}-1$ Hz range) thanks to the large 
collecting area. Variability on such a short timescale would have been 
missed if observed with any previous X--ray satellite (Campana et al. 2003).
In the EPIC-pn instrument light curve three flare-like events can be identified.
Flare activity has been recently reported also in the optical for a number of
transient black holes during quiescence as well as for Cen X-4 (Zurita et
al. 2003, Hynes et al. 2002). 
Flares occur on timescales of minutes to a few hours, with no dependence on orbital 
phase. The mean duration of optical flares in Cen X-4 is 21 min. 
This is similar to what observed in the X--rays.

Small spectral variations are observed as well. In order to highlight the cause of 
this variability, we divided the data into intensity intervals and fit the resulting 
spectra with the canonical model for neutron star transients in quiescence. The variability 
can be mainly accounted for by a variation in the column density together with another 
spectral parameter (either power law index or neutron star atmosphere temperature). 
Based on the available spectra we cannot prefer a variation of the power law 
versus a variation in the temperature of the atmosphere component (even if the
first is slightly better in terms of reduced $\chi^2$, Campana et al. 2003). 
Variations in the neutron star atmosphere might suggest that accretion onto the neutron 
star surface is occuring in quiescence (e.g. Rutledge et al. 2002); variations in the power 
law tail should support the view of an active millisecond radio pulsar emitting X-rays 
at the shock between a radio pulsar wind and inflowing matter from the companion star 
(e.g. Campana \& Stella 2003).

\section{Summary}

SXRT sources are discolsing their secrets thanks to the new astronomical 
facilities. This is mainly
occurring in the X--ray band, as expected, since they emit the bulk of their
electromagnetic radiation in this band. XMM-Newton and Chandra are observing
with unprecedent details  these sources, finding unexpected results also for
well known sources. Observations in other bands (mainly radio and optical) are
providing valuable information too. Two models have been put forward to
explain the main emission properties of SXRTs in quiescence: one involve
accretion onto the neutron star (onto the surface or associated with an ADAF
and a propeller) and the other an active radio pulsar the relativistic wind
of which interacts with the mass outflowing  from the companion. These
emission mechanisms are not encompassed by persistent sources. The ADAF
scenario involves a large mass inflow rate variation from outburst to
quiescence (which is not observed in the parent population of TBHs,
(Campana \& Stella 2000) and a working propeller or a strong outflow. The
pulsar scenario does not involve any accretion onto the neutron star surface
and the luminosity that we see comes from cooling of the neutron star (soft
component) and interation of the relativistic pulsar wind with matter (hard
component).

%%%%%%%%%%%%%%%%%%%%%%%%%%%%%%%%%%%%%%%%%%%%%%%%
%% BACKMATTER
%%%%%%%%%%%%%%%%%%%%%%%%%%%%%%%%%%%%%%%%%%%%%%%%

%\begin{theacknowledgments}
%\end{theacknowledgments}

%%%%%%%%%%%%%%%%%%%%%%%%%%%%%%%%%%%%%%%%%%%%%%%%
%% You may have to change the BibTeX style below, depending on your
%% setup or preferences.
%%
%% If the bibliography is produced without BibTeX comment out the
%% following lines and see the aipguide.pdf for further information.
%%
%% For The AIP proceedings layouts use either
%%%%%%%%%%%%%%%%%%%%%%%%%%%%%%%%%%%%%%%%%%%%

\bibliographystyle{aipproc}   % if natbib is available
%\bibliographystyle{aipprocl} % if natbib is missing

%%%%%%%%%%%%%%%%%%%%%%%%%%%%%%%%%%%%%%%%%%%
%% You probably want to use your own bibtex database here
%%%%%%%%%%%%%%%%%%%%%%%%%%%%%%%%%%%%%%%%%%%
%\bibliography{sample}

\begin{thebibliography}{99}
\bibitem{brown1998} Brown, E. F., Bildsten, L., Rutledge,
R. E.\ 1998, ApJ, 504, L95 
\bibitem{burderi2003}  Burderi, L., Di Salvo, T.,
D'Antona, F., Robba, N.~R., Testa, V.\ 2003, A\&A, 404, L43
\bibitem{burgay2003} Burgay, M., Burderi, L., Possenti,
A., D'Amico, N., Manchester, R.~N., Lyne, A.~G., Camilo, F., Campana, S.\
2003, ApJ, 589, 902  
\bibitem{campana2001} Campana, S. 2001, in ``X--ray astronomy :
stellar endpoints, AGN, and the diffuse X--ray background'', eds. N.E. White,
G. Malaguti, G.G.C. Palumbo, AIP 599 63  
\bibitem{campanastella2000} Campana, S., Stella, L.\
2000, ApJ, 541, 849 
\bibitem{campanastella2003} Campana, S., Stella, L.\
2003, ApJ, in press (astro-ph/0307218) 
\bibitem{campana1997} Campana, S., Mereghetti, S.,
Stella, L., Colpi, M.\ 1997, A\&A, 324, 941 
\bibitem{campana1998} Campana, S., Colpi, M.,
Mereghetti, S., Stella, L., Tavani, M.\ 1998a, A\&A Rev., 8, 279   
\bibitem{campana1998b} Campana, S., Stella, L.,
Mereghetti, S., Colpi, M., Tavani, M., Ricci, D., Dal Fiume, D., Belloni, T.\ 1998b, ApJ, 499, L65 
\bibitem{campana2002} Campana, S., Stella, L.,
Gastaldello, F., Mereghetti, S., Colpi, M., Israel, G.~L., Burderi, L., Di
Salvo, T., Robba, R.~N.\ 2002, ApJ, 575, L15  
\bibitem{campana2003} Campana, S., Israel, G. L., Stella,
L., Gastaldello, F., Mereghetti, S. \ 2004, ApJ in press (astro-ph/0309775) 
\bibitem{colpi2001} Colpi, M., Geppert, U., Page, D.,
Possenti, A. 2001, ApJ, 548, L175 
\bibitem{damico2001} D'Amico, N., et al. 2001, ApJ, 561, L89
\bibitem{fender2003} Fender, R.~P., Gallo, E., Jonker,
P.~G.\ 2003, MNARS, 343, L99 
\bibitem{garcia2001} Garcia, M. R., McClintock, J. E.,
Narayan, R., Callanan, P., Barret, D., Murray, S. S.\ 2001, ApJ, 553, L47 
\bibitem{hynes2002} Hynes, R.~I., Zurita, C., Haswell,
C.~A., Casares, J., Charles, P.~A., Pavlenko, E.~P., Shugarov, S.~Yu., Lott,
D.~A.\ 2002, MNRAS, 330, 1009 
\bibitem{illarionov1975} Illarionov, A.~F.,
Sunyaev, R.~A.\ 1975, A\&A, 39, 185 
\bibitem{kong2002} Kong, A. K.~H., McClintock, J. E.,
Garcia, M. R., Murray, S. S., Barret, D. 2002, ApJ, 570, 277 
\bibitem{menou1999} Menou, K., Esin, A. A., Narayan, R.,
Garcia, M. R., Lasota, J.-P., McClintock, J. E.\ 1999, ApJ, 520, 276 
\bibitem{menou2001} Menou, K., McClintock, J. E.\
2001, ApJ, 557, 304 
\bibitem{pooley2003} Pooley, D., et al. 2003, ApJ, 591,
L131 
\bibitem{rutledge2001} Rutledge, R. E., Bildsten, L.,
Brown, E. F., Pavlov, G. G., Zavlin, V. E.\ 2001, ApJ 551, 921 
\bibitem{rutledge2002}  Rutledge, R. E., Bildsten, L.,
Brown, E. F., Pavlov, G. G., Zavlin, V. E.\ 2002, ApJ, 577, 346
\bibitem{tavani1997} Tavani, M., Arons, J. 1997, ApJ,
477, 439 
\bibitem{yu1996} Yi, I., Narayan, R., Barret, D., McClintock,
J.~E.\ 1996, A\&AS, 120, 187 
\bibitem{wijnands2003} Wijnands, R., Heinke, C. O.,
Pooley, D., Edmonds, P. D., Lewin, W. H. G., Grindlay, J. E., Jonker, P. G.,
Miller, J. M. 2003, ApJ submitted (astro-ph/0310144)
\bibitem{wijnands1998} Wijnands, R., van der Klis, M.\
1998, Nature, 394, 344  
\bibitem{zampieri1995} Zampieri, L., Turolla, R., Zane,
S., Treves, A. 1995, ApJ, 439, 849 
\bibitem{zurita2003} Zurita, C., Casares, J., Shahbaz, T.\
2003, ApJ, 582, 369 

\end{thebibliography}

%%%%%%%%%%%%%%%%%%%%%%%%%%%%%%%%%%%%%%%%%%%
%% Just a reminder that you may have to run bibtex
%% All of it up to \end{document} can be removed
%% if you don't like the warning.
%%%%%%%%%%%%%%%%%%%%%%%%%%%%%%%%%%%%%%%%%%%
%\IfFileExists{\jobname.bbl}{}
% {\typeout{}
%  \typeout{******************************************}
%  \typeout{** Please run "bibtex \jobname" to optain}
%  \typeout{** the bibliography and then re-run LaTeX}
%  \typeout{** twice to fix the references!}
%  \typeout{******************************************}
%  \typeout{}
% }

\end{document}